# Controlled generation of a pn-junction in a waveguide integrated graphene photodetector

Simone Schuler[1,*], Daniel Schall[2], Daniel Neumaier[2], Lukas Dobusch[1], Ole Bethge[3], Benedikt Schwarz[3], Michael Krall[1], and Thomas Mueller[1,*]

[1] *Vienna University of Technology, Institute of Photonics, Gußhausstraße 27-29, 1040 Vienna, Austria*
[2] *AMO GmbH, Otto-Blumenthal-Straße 25, 52074 Aachen, Germany*
[3] *Vienna University of Technology, Institute of Solid State Electronics, Floragasse 7, 1040 Vienna, Austria*

*Corresponding authors: simone.schuler@tuwien.ac.at, thomas.mueller@tuwien.ac.at

**With its electrically tunable light absorption and ultrafast photoresponse, graphene is a promising candidate for high-speed chip-integrated photonics. The generation mechanisms of photosignals in graphene photodetectors have been studied extensively in the past years. However, the knowledge about efficient light conversion at graphene pn-junctions has not yet been translated into high-performance devices. Here, we present a graphene photodetector integrated on a silicon slot-waveguide, acting as a dual-gate to create a pn-junction in the optical absorption region of the device. While at zero bias the photo-thermoelectric effect is the dominant conversion process, an additional photoconductive contribution is identified in a biased configuration. Extrinsic responsivities of 35 mA/W, or 3.5 V/W, at zero bias and 76 mA/W at 300 mV bias voltage are achieved. The device exhibits a 3 dB-bandwidth of 65 GHz, which is the highest value reported for a graphene-based photodetector.**

**Keywords:** graphene, photodetector, photo-thermoelectric effect, integrated photonics

The dense integration of photonic components on a silicon (Si) chip allows for a dramatic increase of the performance of optical communication systems at reduced cost. Photodetectors convert light into electrical signals and are at the heart of any optical link. In silicon photonics, traditionally germanium (Ge) [1,2] or III-V compound semiconductors [3,4] are used for photodetection and both technologies have reached a high level of maturity. Nevertheless, the direct monolithic integration of III-V photodetectors on Si wafers remains a challenge because of the large lattice constant mismatch and different thermal coefficients. Ge can be directly grown on crystalline Si, but pushing the bandwidth of Ge photodetectors to higher and higher frequencies [5,6,7] becomes increasingly difficult because of the material's poor electrical quality. One of the most promising routes to a new era of chip performance is the monolithic 3D integration of electronic and photonic components on the same chip, where a promising material for the photonic layers is SiN [8]. On these amorphous SiN layers, crystalline Ge (the prerequisite for realizing high performance photodetectors) cannot be grown, leaving graphene as the currently only SiN-compatible material that has the potential to enable high-speed photodetection [9].

Recently, graphene [10] has emerged as an attractive material in photonics because of its ultra-broadband light absorption [11,12,13], high carrier mobility [14], and gate-tunability of the optical absorption [15,16,17,18,19]. Moreover, it can be integrated onto virtually any waveguide material, including Si, SiN or AlN. Already early studies have revealed an ultrafast photoresponse in graphene with an intrinsic bandwidth of 260 GHz [20,21,22], showing the promise of graphene for high-speed photodetection applications. Since then, several graphene-based photodetectors on Si waveguides [23,24,25,26,27] have been realized. Devices with a 3-dB cutoff at 42 GHz [26] and detection of a 50 Gbit/s data stream [27] have been demonstrated, on par with state-of-the-art Ge detectors, but the performance of graphene based detectors suffer from a limited controllability of the Fermi level and thus of the photoresponse. In order to further improve graphene-based photodetectors especially in terms of efficiency, improved device concepts are necessary which allow exploiting graphene's remarkable properties.

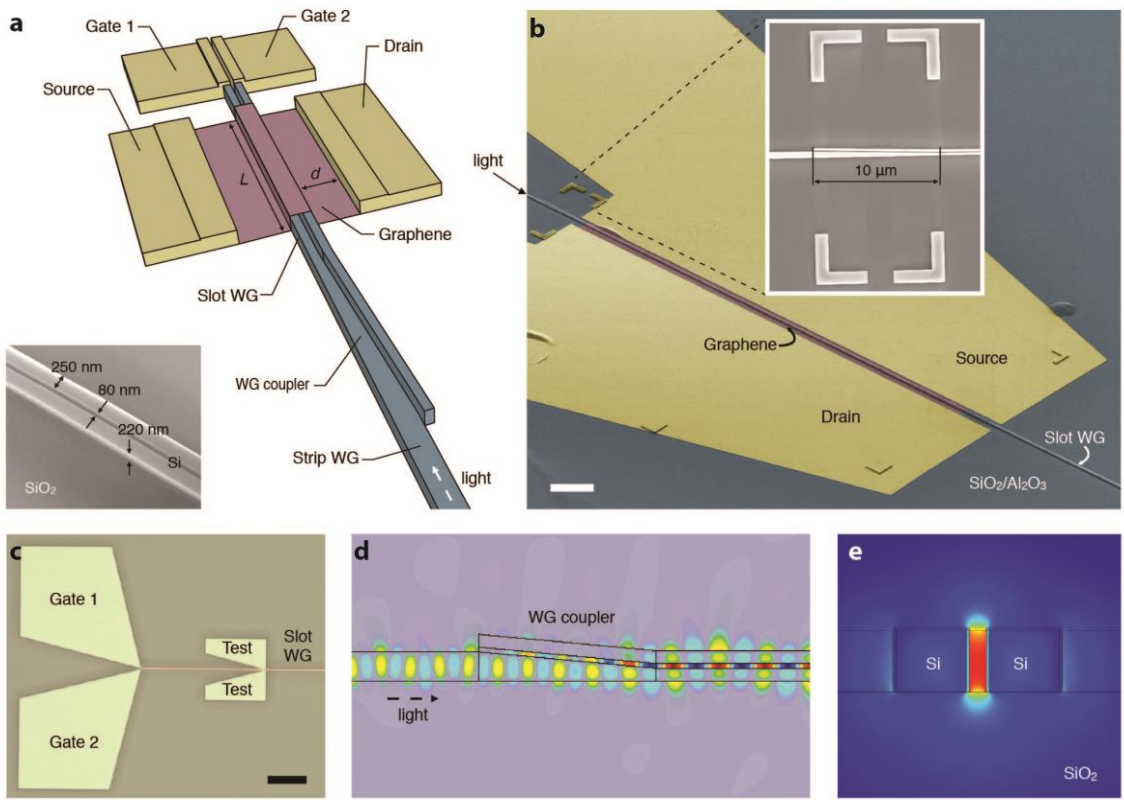

**Figure 1. (a)** Sketch of the graphene photodetector based on a slot-waveguide. Inset: SEM image and dimensions of the slot waveguide. **(b)** Colored SEM picture of a device on Al$_2$O$_3$ and mode coupler (inset). **(c)** Contact pads for the Si slabs, which are used as gate electrodes. Scale bar, 12 μm. **(d)** FEM simulation result for the mode coupler, which couples light from a strip waveguide into the slot-waveguide. **(e)** Electric field distribution of the TE-mode in the slot-waveguide.

In this letter, we present a Si slot-waveguide integrated graphene photodetector relying on the photo-thermoelectric (PTE) effect. The PTE effect plays an important role in graphene-based photodetectors because of the large Seebeck coefficient [28]. In addition the large optical phonon energy (~0.2 eV) [29] and the low scattering rate via acoustic phonons [30,31] give rise to an increased temperature of the photoexcited carriers for some picoseconds, while the lattice stays close to room temperature. As the electronic response is generated from hot electrons, large bandwidths can be achieved [32]. A photovoltage is generated from hot carriers, if the Seebeck coefficient, governed by the doping, as well as the temperature varies in the graphene sheet. The PTE effect [28] has been shown to be dominant in monolayer-bilayer graphene junctions [33], metal-graphene interfaces [34,35,36,37] and partially suspended graphene [38] without external bias. Under bias, photoconductive and bolometric effects also contribute to the photoresponse [39].

The slot-waveguide geometry employed in this work has a twofold function. First, the two silicon strips of the slot-waveguide are used as local gate electrodes to create a controlled and tunable pn-junction in the graphene absorption region. These static gates do not contribute to the radio-frequency (RF) capacitance and therefore leave the bandwidth of the photodetector unaffected. Second, the light is strongly confined in the slot and is hence absorbed by the graphene exactly in between the p-doped and n-doped regions, maximizing the photoresponse due to the PTE effect. A sketch of the device concept is shown in Figure 1(a). We used finite element (FEM) simulations to design the passive photonic structures. The waveguide consists of two strips of a high refractive index material (Si), separated by a subwavelength low refractive (air) slot [40]. The optical mode profile in the slot-waveguide is depicted in Figure 1(e). To efficiently couple light from an optical fiber, the light is first coupled via a grating coupler into a conventional strip waveguide and then adiabatically transferred with high efficiency into the slot-waveguide via a mode converter (Figure 1(d)). Additional information on the grating coupler and the mode converter are provided in the Supporting Information. The waveguides were fabricated on a silicon-on-insulator (SOI) wafer with 220 nm thick Si device layer and 3 μm buried oxide. $L_G$ = 250 nm wide Si strips with a gap of $L_S$ = 80 nm were defined using electron-beam lithography and reactive ion etching. Scanning electron microscopy (SEM) pictures of the waveguide and the mode converter are shown in the insets of Figures 1(a) and 1(b), respectively. Electrical contacts with $L$ = 4 μm drain-source separation were fabricated using electron-beam lithography and metallization (5 nm titanium (Ti), 60 nm gold (Au)). The waveguides were contacted with Ti/Au pads, with additional test pads to verify the ohmic contacts to the Si strips (Figure 1(c)). The silicon waveguides were p-type doped with a resistivity of 14–22 Ωcm.

To prevent electrical contact between the graphene and the waveguide strips, either a $d_G$ = 10 nm thick layer of alumina ($Al_2O_3$) was grown by means of atomic layer deposition (ALD) or hexagonal boron nitride (hBN) of similar thickness was transferred onto the Si waveguide structures. Graphene of proper size and monolayer thickness was prepared by mechanical exfoliation on a stack of polymers on a sacrificial Si wafer. The polymer stack consisted of PAA (poly acrylic acid) and PMMA (poly methyl methacrylate acid) with a thickness chosen such that flakes could be identified by optical microscopy. The monolayer thickness of the graphene was verified using Raman spectroscopy [41] (see Supporting Information). The PAA-PMMA stack was then put into water to dissolve the PAA, thus the PMMA layer with the graphene on top was released from the Si wafer. Afterwards, the PMMA film was positioned on a PDMS (poly-dimethyl-siloxane) stamp, turned upside down and placed with micrometer precision over the slot-waveguide. Care was taken to avoid additional placement of graphitic chunks on top of the photonic structures. Therefore, an aperture was defined from PMMA in a previous step. The PMMA carrier substrate, used for the transfer, was simultaneously used to fabricate electrical contacts to the graphene via electron-beam lithography. Critical point drying was employed to avoid damage of the suspended graphene sheet over the slot. In Figure 1(b) a (colored) SEM image of a typical device on $Al_2O_3$ is shown.

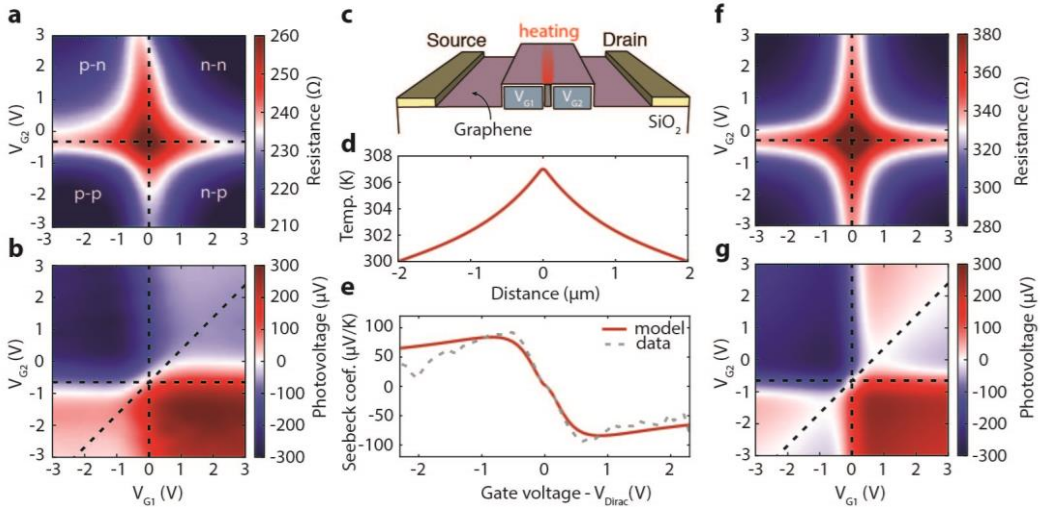

**Figure 2.** Results for Sample A. **(a)** Resistance map for varying gate voltages $V_{G1}$ and $V_{G2}$ applied to the slot-waveguide. Four characteristic regions can be identified; n-n, n-p, p-p and p-n, indicating the gate-tunability of the carriers. The resistance peak indicates the charge neutrality point (Dirac point) of the device. **(b)** Measured photovoltage map at zero bias. **(c)** Sketch of the modeled structure. **(d)** Calculated electron temperature profile over the distance between the source and drain contacts. **(e)** Seebeck coefficient as a function of gate voltage ($V_{G1} = V_{G2} = V_G$) as calculated from the model and the measured data, plotted as solid and dashed lines, respectively. **(f)** Calculated resistance map depending on the gate voltages. **(g)** Corresponding photovoltage map, calculated using the Mott formula. The measured and calculated photovoltage are in good agreement.

For electrical and optical characterization, the two silicon gates were wire-bonded while drain and source were contacted using an RF probe in signal-ground (SG) configuration. The devices were characterized electrically by varying the two gate voltages, $V_{G1}$ and $V_{G2}$, at a fixed drain-source voltage $V_{DS}$ and recording the device current $I_{DS}$. The resulting resistance map of a $W = 30$ µm long monolayer device on $d_G = 13$ nm of hBN (sample A) is shown in Figure 2(a). Four characteristic regions can be identified: n-n, n-p, p-p and p-n, which indicate the gate tunability of the carrier density in the graphene sheet. For electro-optical characterization we used chopped light from a telecom laser diode with a wavelength of 1560 nm ($E_{ph} \approx 0.8$ eV photon energy), which was coupled into an optical fiber and further coupled via the grating coupler into the chip. While varying the gate voltages $V_{G1}$ and $V_{G2}$, the photovoltage was recorded using a lock-in amplifier. The resulting photovoltage map, presented in Figure 2(b), shows the typical six-fold pattern that serves as a fingerprint for the PTE effect [28,35] (see also Supporting Information).

As our data suggest the PTE effect to be the dominant conversion process, we performed simulations based on the model depicted in Figure 2(c). In general, the photovoltage generated from the PTE effect can be calculated by integrating the optically induced temperature gradient $\nabla T_e$, with the locally varying Seebeck coefficient, $V_{PTE} = \int S(x) \, \nabla T_e(x) \, dx$. The Seebeck coefficient $S$ is related to the electrical conductivity $\sigma$ via the Mott equation, $S = -\pi^2 k_B^2 T_e/(3q) \, \sigma^{-1} \, \partial\sigma/\partial\mathcal{E}$, where $q$ is the electron charge, $k_B$ denotes the Boltzmann constant and $\mathcal{E}$ the Fermi energy [42]. To obtain an analytical expression for the Seebeck coefficient for further modeling, we approximate the graphene conductivity by $\sigma = \sqrt{\sigma_{min}^2 + [\mu C_G (V_G - V_{Dirac})]^2}$, where $\sigma_{min}$ denotes the minimum conductivity, $\mu$ is the carrier mobility, $C_G = \epsilon_G/d_G$ is the gate capacitance and $V_{Dirac}$ the charge neutrality point voltage. Care has to be taken to include the contact resistance $R_C$ and the excess resistance $R_U$ of the ungated regions to both sides of the waveguide. The overall device conductance can be written

$$G = \frac{1}{R_C + R_U + (L/W) \, \sigma^{-1}} \, .$$

Setting $R_C + R_U = 254$ Ω results in the expected linear relation between $\sigma$ and $V_G$, as shown in the Supporting Information. From this we can then estimate the intrinsic graphene conductivity and obtain $\sigma_{min} \approx 0.23$ mS, $\mu \approx 2000$ cm²/Vs and $V_{Dirac} \approx 0.7$ V. Figure 2(f) depicts the calculated resistance map as a function of the two gate voltages.

Based on these values, the Seebeck coefficient was calculated. The result is illustrated in Figure 2(e). The solid line depicts the Seebeck coefficient obtained from the analytical σ-model, used for further simulations, and the dashed line shows the Seebeck coefficient directly calculated from the experimental data. The electron temperature $T_e$ is obtained by solving the heat equation [26]

$$\frac{\partial^2 T_e}{\partial x^2} + \frac{\alpha P'}{\kappa_e} - \frac{T_e - T_l}{L_C^2} = 0$$

for the given geometry, where $\alpha P'$ denotes the absorbed power density in the slot, $P' = P_{in}/(L_S W)$, $P_{in}$ is the incident optical power in the waveguide, $L_C \approx 1$ μm [37] is the characteristic cooling length in graphene, and $T_l$ = 300 K the lattice temperature. $\kappa_e$ is the electronic thermal conductivity that can be determined from the Wiedemann-Franz relation $\kappa_e = \sigma L_0 T$ [42], where, for simplicity, we use the standard Lorenz number $L_0 = 2.4 \times 10^{-8}$ (V/K)$^2$ and set $T$ to 300 K. In order to estimate the absorption $\alpha$, we performed FEM simulations (see Supporting Information). The calculated temperature profile is shown in Figure 2(d). We assumed the Seebeck coefficient to be constant over the gated waveguide regions and zero in the ungated regions. The resulting photovoltage map, presented in Figure 2(g), is in good agreement with the measurement, indicating that the PTE effect is indeed the dominant mechanism.

Given an incident optical power in the fiber of $P_{fiber} \approx 0.8$ mW, a $\eta_C \approx 0.4$ coupling efficiency of the grating coupler (see Supporting Information), and assuming no losses in the 2 mm long waveguides and the mode coupler, we estimate a photoresponsivity of $R_V = V_{PTE}/(\eta_C P_{fiber}) \approx 0.9$ V/W under zero-bias operation. We also conducted independent measurements of the photocurrent $I_{PTE}$ under same conditions using a current amplifier with low input impedance (the corresponding photocurrent map is presented in the Supporting Information), which resulted in a responsivity of $R_I \approx 3$ mA/W. From the relation $V_{PTE} = I_{PTE} R$ we estimate a device resistance of $R \approx 300$ Ω, in agreement with the electrical I-V measurements.

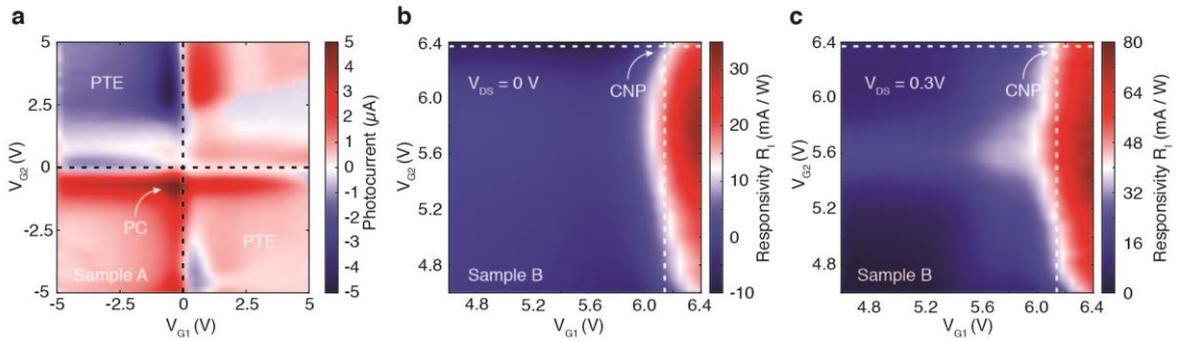

**Figure 3. (a)** Photocurrent map (Sample A) at $V_{DS}$ = -400 mV. **(b)** Responsivity map of our best device (Sample B) at zero bias, where we achieved a peak responsivity of 35 mA/W. CNP; charge neutrality point. **(c)** Responsivity map at $V_{DS}$ = 300 mV (Sample B). The responsivity increases to 76 mA/W.

Figure 3(a) shows a photocurrent map recorded with a drain-source bias of $V_{DS}$ = -400 mV. The photoresponse shows a complex behavior with varying $V_{G1}$ and $V_{G2}$ and is dominated by a cross-shaped feature around the Dirac point (see dashed lines). The maximum responsivity increases to ~18 mA/W. In order to get an understanding of the underlying photocurrent generation mechanism, we first performed simulations of the PTE effect that showed a negligible influence of the drain-source voltage. Previous work [39], however, has demonstrated that bolometric (BL) and photoconductive (PC) effects play an important role under bias. We find the BL effect to be inconsistent with the experimental results, both in sign and gate voltage dependence. The PC effect can be modeled by

$$I_{PC} \approx \left(\frac{L_S}{L}\right) W \Delta \sigma E$$

where the prefactor ($L_S/L$) accounts for the fact that only part of the graphene sheet is optically excited. $E$ denotes the lateral electric field at the waveguide and $\Delta \sigma$ is the photoexcited conductivity. The latter can be expressed as $\Delta \sigma = q\mu(\Delta n_e + \Delta n_h)$, with $\Delta n_e = \Delta n_h \approx \alpha P_{in}\tau_L/(E_{ph}WL_S)$ being the photo-generated electron and hole densities, respectively, and $\tau_L \approx$ 2 ps the carrier lifetime in graphene [20]. It is instructive to consider first the case of a homogeneous graphene channel (i.e. $V_{G1} \sim V_{G2} \sim$ 0) and estimate the expected PC current magnitude. At $|V_{DS}|$ = 400 mV, we obtain $E \approx |V_{DS}|/L \approx$ 1 kV/cm and $R_I \approx 10^{-2}$ A/W, in agreement with the experiment (~18 mA/W), which leads us to conclude that the PC effect is indeed dominant. The drop of the PC response with increasing carrier concentration/gate voltage is expected [39] and can be attributed to several factors, such as screening of the lateral electric field, shortening of the carrier lifetime, and enhanced electron-electron scattering. The sign of the photocurrent is (mostly) positive, also consistent with the PC effect.

In an improved set of devices, we replaced the hBN gate dielectric by $Al_2O_3$, with a twofold motivation: (i) SEM pictures (see Supporting Information) revealed that the hBN layer does not conformally cover the waveguide. As a result, only a fraction $\propto L_G/L_C$ of the electron heat is converted into a photovoltage. In devices with $Al_2O_3$, on the contrary, the sidewalls of the waveguide contribute to the gating as well, which approximately doubles $V_{PTE}$. Additionally, less cooling in the freely suspended graphene sheet leads to higher electron temperatures [43]. (ii) The large series resistance $R_U$ in devices with hBN severely reduces $I_{PTE}$, according to $I_{PTE} = V_{PTE}/R$. Using $Al_2O_3$, instead, charge transfer between graphene and the oxide leads to p-doping of the ungated regions, which reduces $R_U$. With these modifications, the overall device resistance dropped to ~110 Ω and the (zero-bias) responsivities increased to $R_V \approx$ 3.5 V/W and $R_I \approx$ 35 mA/W, respectively (sample B; Figure 3(b)). From electrical characterization we extracted a mobility of ~1000 cm²/Vs (see Supporting Information). Although $V_{Dirac}$ shifted to ~ 6.2 V, the high-responsivity p-n and n-p regimes can still be reached by applying gate voltages without breaking the Al₂O₃ dielectric. Under a moderate drain-source voltage of $V_{DS}$ = 300 mV (Figure 3(c)), a peak responsivity of $R_I \approx$ 76 mA/W is achieved.

The electrical bandwidth of a photodetector is a key indicator of its performance. We performed impulse response measurements, where ~1 ps long optical pulses, generated by a mode-locked erbium fiber laser with a wavelength of 1550 nm, were coupled into the device (sample B) and the impulse response was monitored with an oscilloscope (Figure 4(a)). From this measurement we extract a pulse duration at full-width at half-maximum (FWHM) of $\Delta t \approx$ 30 ps. The electrical bandwidth can then be derived using the time-bandwidth product. Assuming a Gaussian impulse response, $f_{3dB} \approx 0.44/\Delta t \approx$ 15 GHz is obtained, which is the limit of our measurement setup. The Fourier-transform of the pulse, shown in the inset of Figure 4(a), yields the same bandwidth.

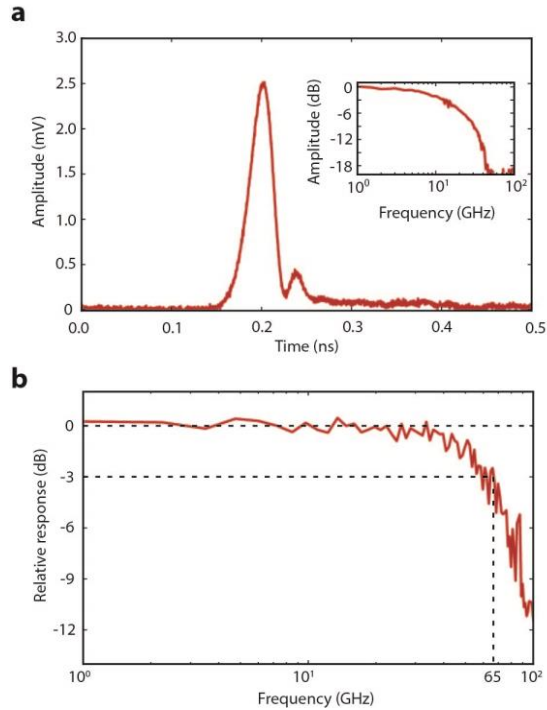

**Figure 4.** (a) Measured impulse response of Sample B using a train of ~1 ps pulses. The pulse duration is $\Delta t \approx$ 30 ps (FWHM), which corresponds to a bandwidth of ~15 GHz, limited by the measurement setup. The Fourier- transformation of the pulse is shown in the inset. (b) Measured frequency response ($V_{DS}$ = 0.5 V) using a heterodyne technique. From this we find a cut-off bandwidth of $f_{3dB} \approx$ 65 GHz.

In order to determine the cut-off frequency of our detector, we used a heterodyne measurement technique similar to the one used in reference [27]. Two laser sources with different frequencies were multiplexed, causing amplitude beating at the difference frequency. Keeping the frequency of one laser constant while tuning the other, the difference frequency can be varied between 1 and 100 GHz. The laser light was amplified with an erbium-doped fiber amplifier and coupled into the graphene photodetector. A 67 GHz SG-probe was used to contact the device and the output power was detected with an RF power meter. A sketch of the measurement setup is depicted in the Supporting Information. The frequency response of the detector is shown in Figure 4(b). From this measurement we obtain a 3-dB cut-off frequency of $f_{3dB} \approx$ 65 GHz, independent of bias voltage, which translates into a potential bit rate of ~90 Gbit/s (for a single wavelength channel and on-off keying), and is the highest value reported for a graphene photodetector. The maximum output power measured at 1 GHz was -31 dBm at 19.4 dBm optical input power and 1.2 V bias and defines the highest RF output power delivered by a graphene detector.

In summary, we have presented an ultrafast graphene-based photodetector on a Si slot-waveguide, where the Si strips serve as local back gate electrodes to create a pn-junction for efficient photodetection. A responsivity of 35 mA/W, or 3.5 V/W, at zero-bias conditions was achieved, while under a moderate drain-source bias of 300 mV, the responsivity increased to 76 mA/W. The photodetector has shown a record high 3-dB cutoff frequency of 65 GHz. To further improve the responsivity in terms of V/W the electrical gating could be extended closer to the contacts, for example, by using thin Si slabs to both sides of the waveguide. Reducing the electrode spacing would result in lower device resistance and thus improved responsivity in A/W.

**Supporting Information:** Design of mode converter and grating coupler, measurement of waveguide and coupling losses, graphene Raman spectrum, scanning electron microscopy images of devices with hBN and $Al_2O_3$ gate dielectrics, details on electrical and optical characterization methods, additional photocurrent and photovoltage maps, extracted intrinsic graphene conductivity, calculated optical absorption in the graphene sheet, sketch of the frequency response measurement setup.

**Acknowledgment:** We thank Andreas Pospischil and Marco Furchi for technical assistance and Frank Koppens, Ilya Goykhman and Marco Romagnoli for helpful discussions. Financial support by the European Union (grant agreement No. 696656 Graphene Flagship) and the Austrian Science Fund FWF (START Y 539-N16) is acknowledged.

**Conflict of Interest:** The authors declare no conflict of interest

# Controlled generation of a pn-junction in a waveguide integrated graphene photodetector

**Simone Schuler**[1], **Daniel Schall**[2], **Daniel Neumaier**[2], **Lukas Dobusch**[1], **Ole Bethge**[3], **Benedikt Schwarz**[3], **Michael Krall**[1], and **Thomas Mueller**[1]

[1] *Vienna University of Technology, Institute of Photonics, Gußhausstraße 27-29, 1040 Vienna, Austria*

[2] *AMO GmbH, Otto-Blumenthal-Straße 25, 52074 Aachen, Germany*

[3] *Vienna University of Technology, Institute of Solid State Electronics, Floragasse 7, 1040 Vienna, Austria*

## 1.) Optical components

*Comsol Multiphysics* was used to perform FEM simulations to design the slot-waveguide as well as the mode converter, which couples the optical mode from a strip waveguide into the slot-waveguide. The guided mode in the strip waveguide couples via the evanescent field into the slot-waveguide. The coupling efficiency dependence on the length of the coupler is illustrated in Figure S1.

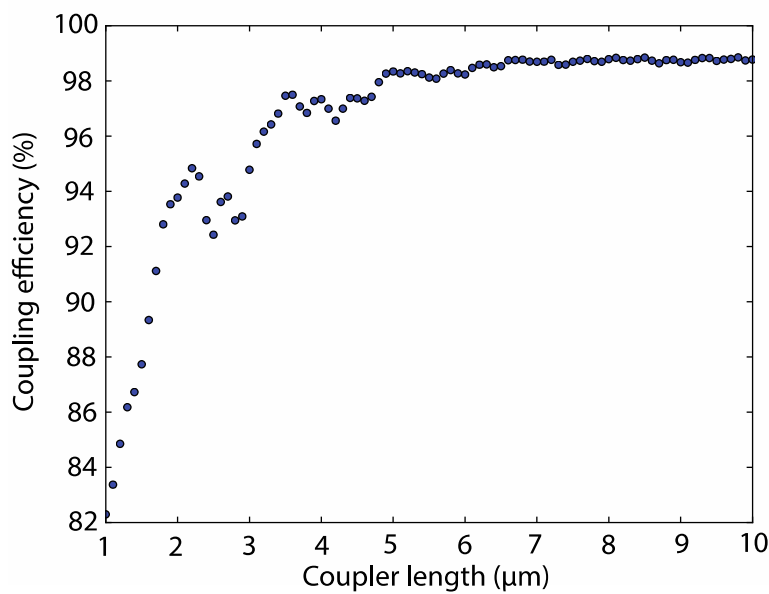

*Figure S1.* Calculated coupling efficiency, depending on the length of the mode converter.

To efficiently couple light from an optical fiber into the strip waveguide, a grating coupler was used as shown in Figure S2a. To determine the responsivity, the losses of the coupler, the waveguide and the mode converter have to be taken into account. Figure S2b shows the dependence of the losses on the length of the waveguide. From the same plot the coupling efficiency of the grating coupler is determined.

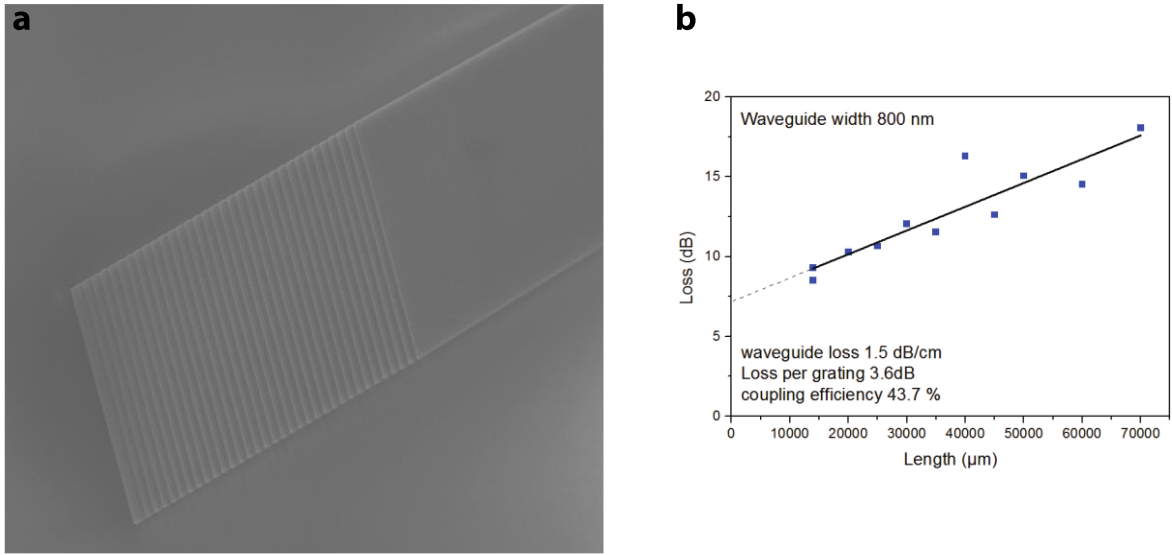

*Figure S2. (a) SEM image of the grating coupler. (b) Waveguide and grating coupler losses.*

## 2.) Device fabrication

The thicknesses of the graphene flakes were defined using Raman spectroscopy. The flakes were characterized on the PAA/PMMA stack before transfer. Figure S3 shows a Raman spectrum of a monolayer of graphene.

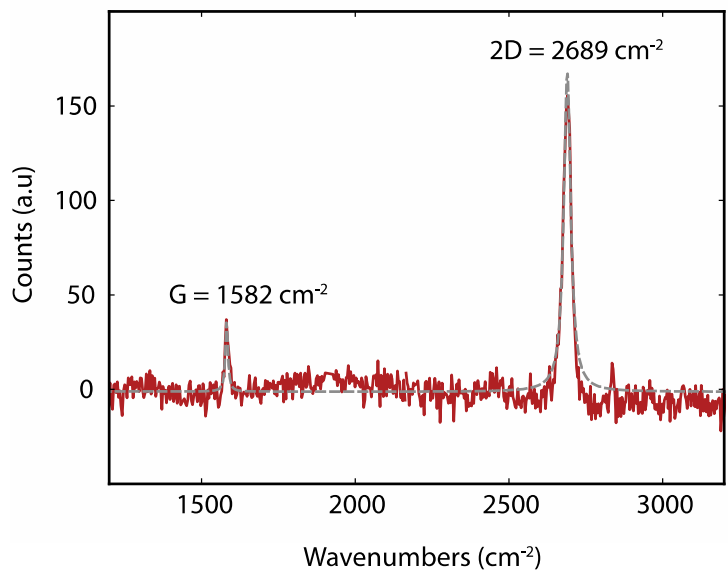

*Figure S3. Raman spectrum of a graphene monolayer on a PAA/PMMA stack.*

Either hexagonal boron nitride or aluminum oxide was used as gate dielectric to prevent electrical contact between the waveguide and graphene. Comparing the performance of devices based either on one or the other, we found a better response in case of aluminum oxide. As discussed in the main text, this is attributed to the non-conformal covering of the waveguide in case of boron nitride. To verify this presumption, we took SEM images of both types of devices. Indeed, the 13 nm boron nitride layer does not conformally cover the waveguide (Figure S4a). Using aluminum oxide as gate dielectric, which is deposited in an ALD-process, the graphene layer covers the waveguide as well as the sidewalls (Figure 4b).

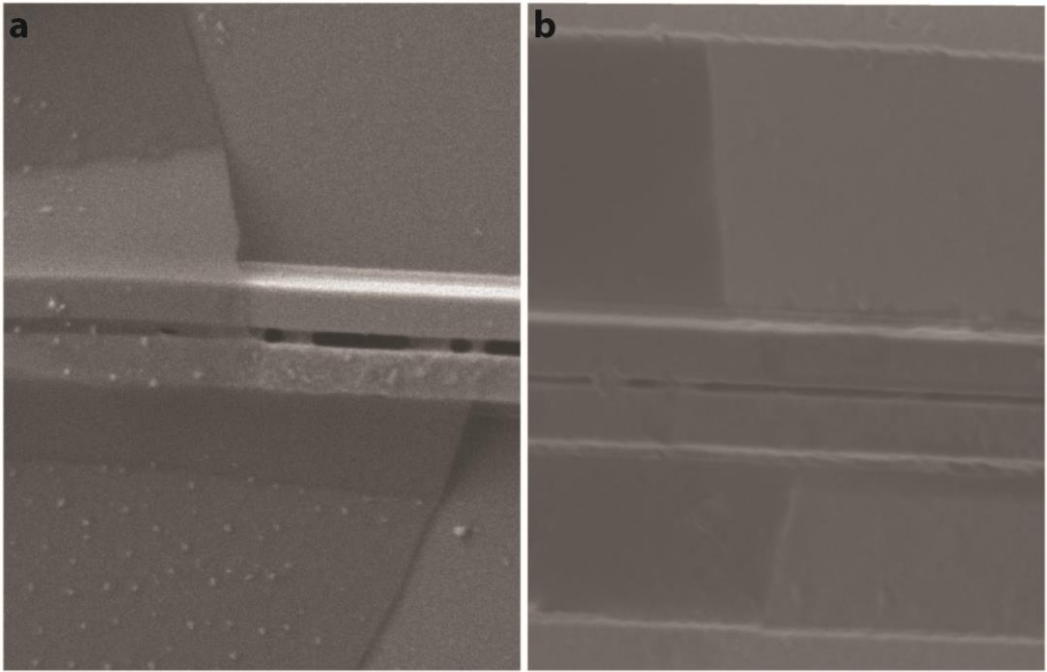

*Figure S4. (a)* *SEM picture of a 13 nm thick hexagonal boron nitride flake on a slot waveguide. (b) SEM image of a graphene flake placed on top a slot waveguide covered with 10 nm of aluminum oxide. As the graphene layer conformally covers the waveguide, the gating length is increased compared to devices with boron nitride.*

### 3.) Electrical characterization

Electrical characterization was performed using a Semiconductor Parameter Analyzer (Agilent 4155C). The devices were contacted using a RF-probe needle in signal-ground configuration (Picoprobe, GGB Industries).

### 4.) Optical characterization

For optical characterization, 1560 nm light from a laserdiode was coupled via polarizers and a lens into an optical fiber (SMF-28), further coupled into the strip waveguide, via the grating coupler, and then into the slot-waveguide using the mode converter. The polarizers were adjusted such that a maximum photocurrent signal was detected. The signal was recorded using a transimpedance amplifier and a lock-in amplifier. We measured both the open-circuit photo*voltage* as well as the short-circuit photo*current* response of our detectors. In Figure S5a the photovoltage map of sample A is presented and Figure S5b shows the corresponding photocurrent map. Our best device (sample B) exhibited a responsivity of 3.5 V/W, respectively 35 mA/W, under zero-bias conditions. Figure 6 shows the resistance map of this device. The photovoltage and photocurrent maps for sample B are presented in Figures S7a and S7b, respectively.

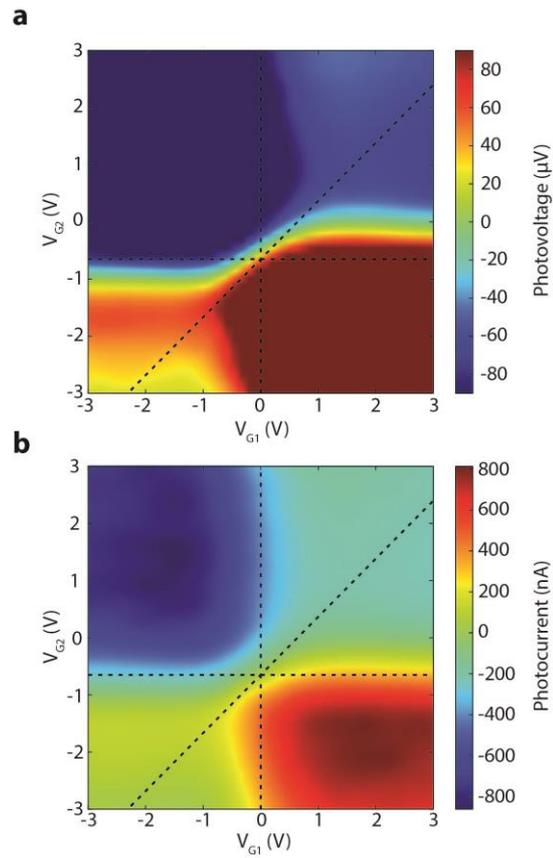

*Figure S5. (a)* Photovoltage map of sample A, illustrating the six-fold photocurrent generation pattern (same Figure as in the main paper, but plotted on a different color scale). *(b)* Corresponding photocurrent map.

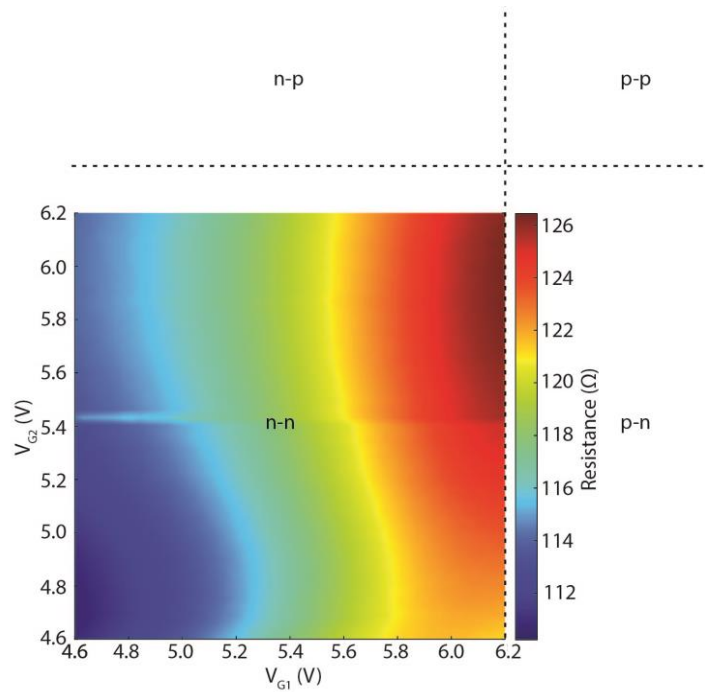

*Figure S6.* Resistance map of sample B.

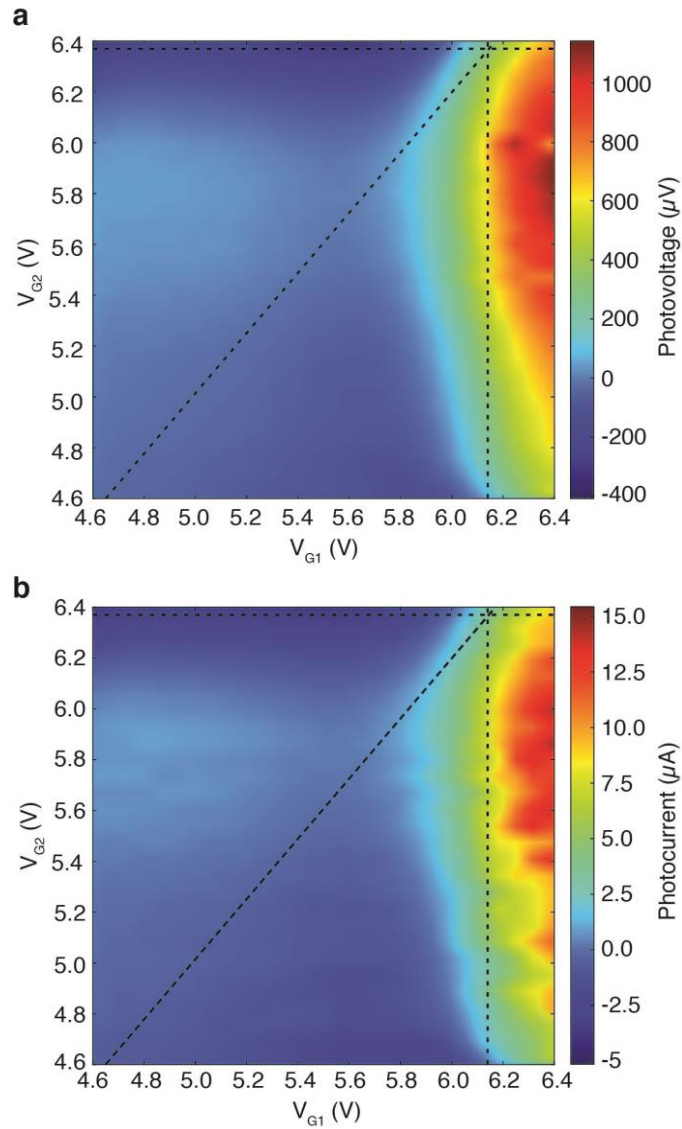

*Figure S7. (a) Photovoltage map of sample B. (b) Photocurrent map.*

### 5.) Modeling of the photo-thermoelectric effect

The photo-thermoelectric effect of sample A was modeled based on the transfer characteristic of the device, as shown in Figure S8a. The Seebeck coefficient was calculated form the intrinsic conductivity, which is shown in Figure S8b. In order to estimate the absorbed power, we used FEM simulation to estimate the absorption depending on the length of the flake for mono- and bi-layer graphene on the slot-waveguide geometry as shown in Figure S9.

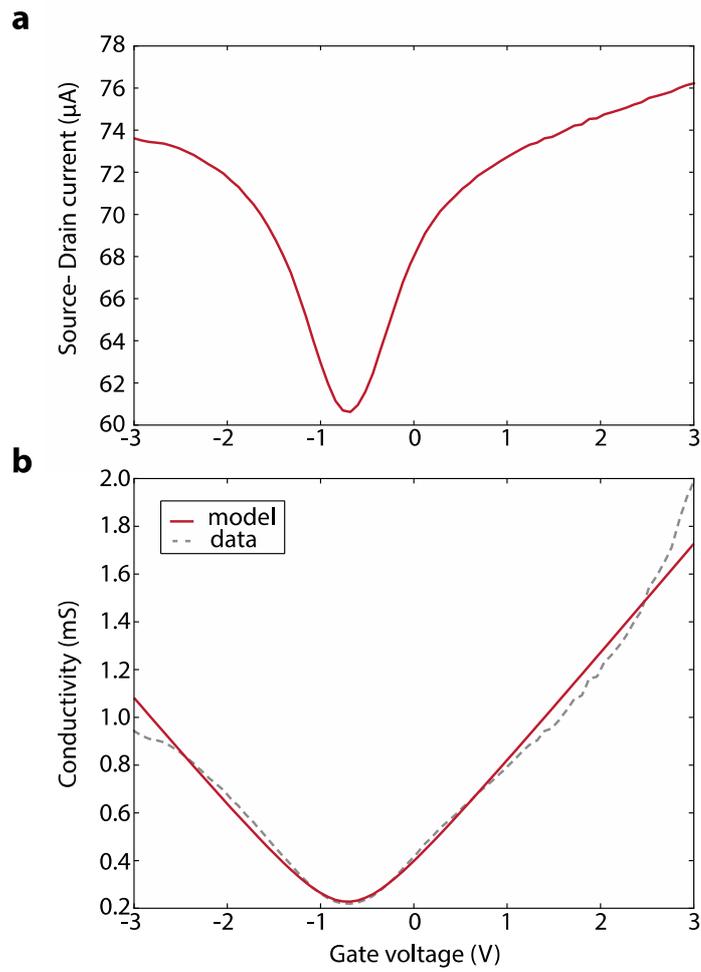

*Figure S8.* (*a*) *Measured transfer characteristic of the presented device with $V_{G1} = V_{G2} = V_G$.* (*b*) *Corresponding intrinsic conductivity and model as dashed and solid lines, respectively.*

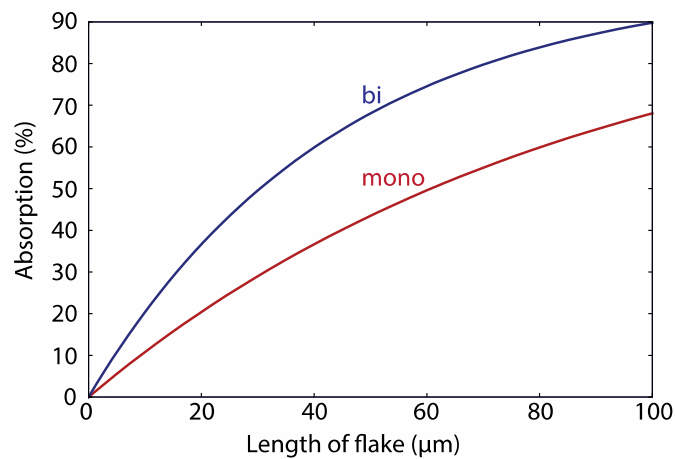

*Figure S9. Calculated absorption coefficient of a mono- and bi-layer flake placed on top of our slot-waveguide geometry, red and blue line, respectively.*

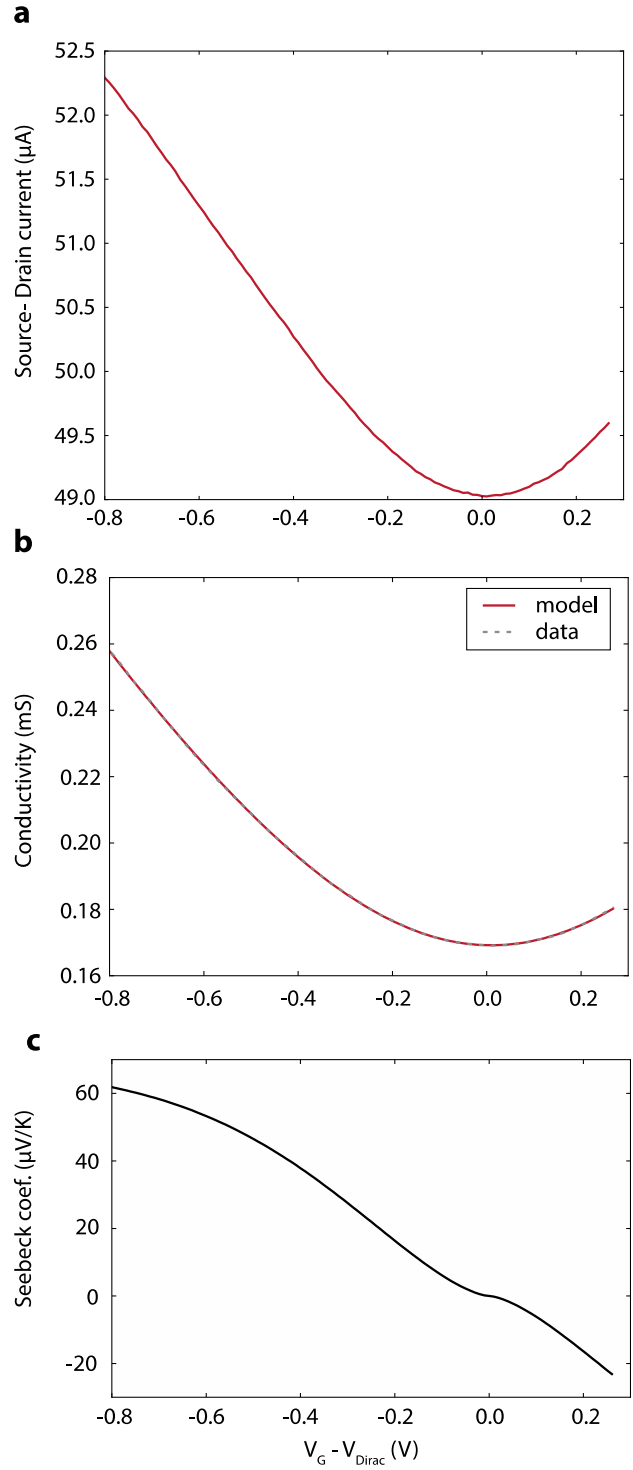

*Figure S10.* (*a*) Measured transfer characteristic of Sample B with $V_{G1} = V_{G2} = V_G$. (*b*) Corresponding intrinsic conductivity data and model, respectively and (*c*) calculated Seebeck coefficient.

## 6.) Photoconductive effect

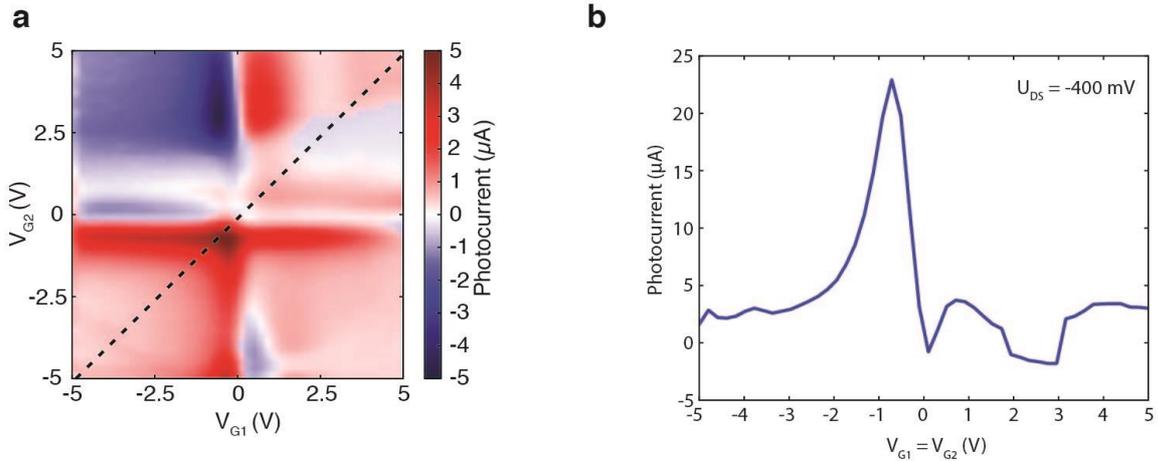

*Figure S11.* (*a*) *Photocurrent map (Sample A) at $V_{DS}$ = -400 mV.* (*b*) *Corresponding photocurrent at the diagonal (dashed line) where the PTE effect is negligible, indicating an additional photoconductive (PC) contribution when a bias is applied.*

## 7.) Frequency response measurement

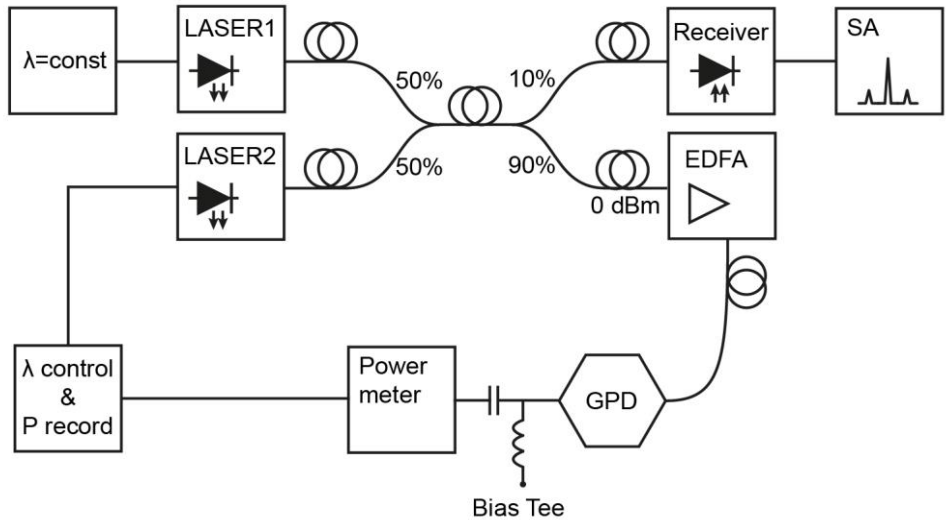

*Figure S12. Sketch of the RF measurement setup. GPD = graphene photodetector, SA = spectrum analyzer, EDFA = erbium doped fiber amplifier. Laser 1 is kept at constant wavelength, the wavelength of Laser 2 is tuned. For each frequency the measured power is recorded. The values in % give the coupling ratio of the fiber couplers.*